# Integration of Computer Networks and Artificial Neural Networks for an AI-based Network Operator

**Binbin Wu[1]\*,Jingyu Xu[2],Yifan Zhang[3], Bo Liu[4],Yulu Gong[5],Jiaxin Huang [6]**

1*Heating Ventilation and Air Conditioning Engineering,Tsinghua University, Beijing China
2 Computer Information Technology,Northern Arizona University,1900 S Knoles Dr, Flagstaff, AZ,USA
3 Executive Master of Business Administration,Amazon Connect Technology Services (Beijing), Co., Ltd.Xi'an, Shaanxi, China
4 Software Engineering,Zhejiang University,HangZhou China
5 Computer & Information Technology,Northern Arizona University,Flagstaff, AZ, USA
6 Information Studies,Trine University,Phoenix USA

\*Corresponding author:[Binbin Wu,E-Mail:wubinbin.1@gmail.com]

**Abstract.**

This paper proposes an integrated approach combining computer networks and artificial neural networks to construct an intelligent network operator, functioning as an AI model. State information from computer networks is transformed into embedded vectors, enabling the operator to efficiently recognize different pieces of information and accurately output appropriate operations for the computer network at each step. The operator has undergone comprehensive testing, achieving a 100% accuracy rate, thus eliminating operational risks. Furthermore, a novel algorithm is proposed to emphasize crucial training losses, aiming to enhance the efficiency of operator training. Additionally, a simple computer network simulator is created and encapsulated into training and testing environment components, enabling automation of the data collection, training, and testing processes. This abstract outlines the core contributions of the paper while highlighting the innovative methodology employed in the development and validation of the AI-based network operator.

## 1. Introduction

The current enterprise network architecture faces numerous challenges during the process of digital transformation. As a critical and complex component of IT infrastructure, computer network systems are undergoing a paradigm shift in architecture and design principles from traditional hardware-centric network operations to business and application-oriented network operations. This transition represents a pivotal goal in the digital transformation of enterprise networks.

Challenges such as insufficient agility in network architecture, inadequate automation and intelligence in network operations, and difficulties in network fault localization are among the primary obstacles. These challenges further result in issues such as failure to meet [1]Service Level Agreement (SLA) requirements for business continuity and high network operation costs.
To address these challenges, this paper introduces the design of an intelligent operator capable of analyzing network information, identifying network faults, and providing appropriate actions to resolve them. This intelligent operator represents a novel approach towards enhancing network resilience, automating network operations, and meeting the evolving demands of digital transformation in enterprise networks.

## 2. Related work

*2.1. Intelligence of computer network system*

Based on network technology, computer network system connects computers, terminals, external devices, servers and other devices distributed in different locations through network devices and communication lines, and realizes information transmission and resource sharing under the control of network communication protocols and network operating systems. An information network is formed by connecting media such as twisted pair wire and optical fiber.

The design of computer network system not only needs to fully meet the existing data transmission, but also needs to meet the development in the next few years. In terms of safety and stability, it is required to have multiple safety and reliability guarantees, and in the design and construction of the equipment, it can provide multiple stability and reliability operation guarantees, and fully ensure the normal operation of the information and intelligent system. [2]The computer network subsystem of intelligent building adopts the star topology structure networking, and generally divides the network architecture into three networks, which are public network, intelligent network, and business private network (such as the medical private network of a hospital, the guest network of a hotel, etc.). When the network boundary firewall is used to isolate the Internet, all network security devices are connected to the core switch in bypass mode to prevent network impact and performance bottlenecks.

The main purpose of establishing computer network is to realize the sharing of computer resources. Computer resources mainly refer to computer hardware, software and data. Networked computers follow a common network protocol.

A stable computer network platform is the foundation of construction, which determines the digitization and information construction level of a city's library, archives and urban construction archives. The demand on the network mainly includes:

1, the construction of a city library, archives, urban construction archives of the external network, internal network and intelligent network: the external network is mainly used as the Internet and internal LAN[3], the external network consider wireless full coverage; The Intranet carries internal application services and connects to the upper-layer Intranet. Intelligent network is an important means to enhance the security level and improve the internal management level and efficiency.

2, the construction of security system: according to the different security level requirements of business and office and relevant policy requirements, combined with the actual situation of the network and application, delimit different security domains. In order to ensure the integrity of the network system, the core switches, access switches and wireless devices of each network are of the same brand.

*2.2. System architecture*
In the example of a minicomputer network in this article, we will show how to make an intelligent operator. This computer network is created randomly through simulators, and in doing so increases the diversity that allows our intelligent operators to make appropriate decisions and reactions in a variety of situations.

First, the basic structure of minicomputer networks. It includes multiple computer nodes, routers, switches, and various network devices. [4]These devices communicate with each other via wired or wireless connections, forming a complex network topology. Our intelligent operators will manage and maintain this network by monitoring network traffic, analyzing packets, detecting network anomalies, and more. It will perform various tasks according to pre-set policies and algorithms, such as:

1. Network monitoring and fault detection: Intelligent operators will regularly scan network traffic, monitor equipment operating status, and detect any abnormal conditions, such as network congestion, equipment failure, etc.

2. Traffic management: It can dynamically adjust the network according to the traffic load and optimize the data transmission path to ensure the efficient operation of the network.

3. Security protection: Intelligent operators will monitor network security vulnerabilities and attacks in real time and take necessary measures to protect network security, such as intrusion detection, firewall configuration, etc.

4. Automated tasks: It can perform a variety of automated tasks, such as automatically backing up data, regularly updating software patches, configuring network devices, etc., to reduce the workload of administrators.

5. Performance optimization: Intelligent operators will provide optimization suggestions and adjustment plans based on network usage and performance indicators to improve the overall performance and stability of the network[5].

With these capabilities, intelligent operators will be able to effectively manage and maintain this small computer network, improving network reliability, security, and performance, thereby providing users with a better network experience. At the same time, it can reduce the workload of administrators and make network management more intelligent and efficient.

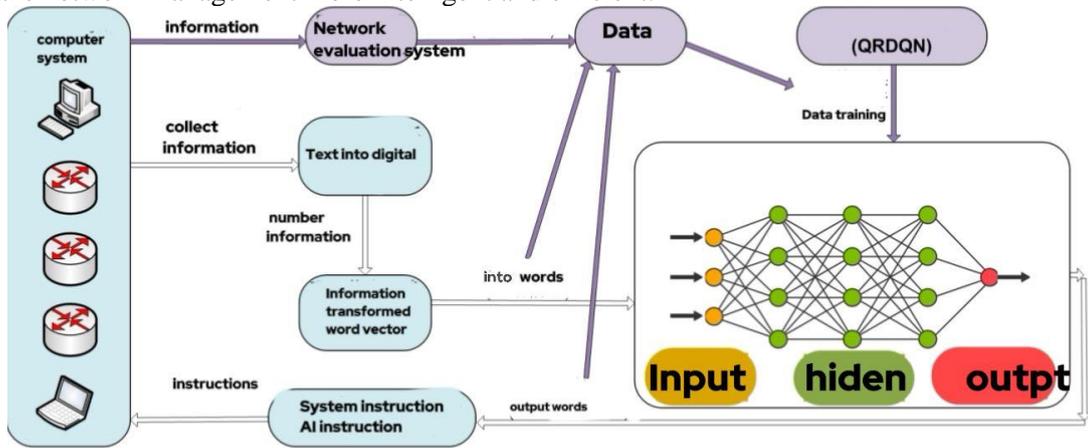

**Figure 1.** Architecture and data flow diagram

A brief explanation of the parts in Figure 1:

Computer network: It can be normal or faulty, and its status information is expressed in text form. It can also update its own status information accordingly according to network device instructions.

Network information: Information about the operating status of a computer network.

Network information (observed information) : [6]The portion of network information that is preprocessed and fed into a neural network for analysis.

An encoder that observes information text into numbers: It encodes information in text form into information in digital form using a predefined mapping table.

Digital (observed information) : Information in digital form that represents the same information as network information (observed information).

An encoder that observes the conversion of information into embedded vectors: It converts numbers into embedded vectors, which are represented by floating-point numbers (vectors). It uses predefined rules for transformation.

Embedded vector (observation information) : floating-point numbers (vectors) that represent the same information as numbers (observation information).

Intelligent operator (artificial neural network model) : It can compute embedded vectors (observation information) and output numbers (instructions that the AI outputs). It is a neural network model that is initially randomly initialized and trained to become an intelligent operator capable of analyzing embedded vectors (observation information) and repairing faults.

Numbers (AI output commands) : Numbers representing network device commands.

Codecs for network device instructions and AI output instructions: It decodes digital (AI output instructions) into network device instructions, or encodes network device instructions into digital (AI output instructions). It uses a predefined mapping table for the transformation.

Network device instruction: It is an operation instruction used to fix the fault, or the conclusion that there is no fault in the embedded vector (observation information).

Network performance evaluation system: It analyzes network information, evaluates the changes of the computer network after the use of network equipment instructions, and then outputs a number (reward point value).

Numbers (reward points) : positive or negative numbers that represent whether the change is good or bad.

Data set: It is a collection of data generated in each loop of observation information/output instruction, including current observation information, current action instruction, current reward, and next observation information.

Reinforcement Learning Algorithm [7](QRDQN) : It uses specific algorithms and data sets to train operators, this article uses the QRDQN algorithm and the critical loss algorithm.

*2.3. Intelligent operator*

As a neural network model, the training of intelligent operator requires a large amount of real-time data and feedback, which needs to be collected from the computer network in real time. To achieve the goal of automating data collection and operator training, we need to integrate computer networks and neural networks together[8].

For example, let's consider a practical network operation scenario: network traffic management. In this case, intelligent operators need to monitor network traffic in real time, identify and analyze different types of packets, and make dynamic adjustments based on network traffic load to optimize network performance. To achieve this, we can take the following steps to integrate computer networks and neural networks:

1. Data acquisition and preprocessing: First, we need to deploy data collectors in the network to capture network traffic data. This data can include information about incoming and outgoing packets, traffic load, network latency, and so on. These original data need to be pre-processed, including data cleaning, feature extraction, etc., in order to facilitate the processing of neural network models.

2. Neural network model design: Next, we design a neural network model suitable for network traffic management tasks. This model can be a deep reinforcement learning model, such as [8-9]Deep Q networks (DQN) or policy gradient methods, for learning optimal policies for network traffic management. The input of the model can be the pre-processed network traffic data, and the output is the corresponding operational decisions, such as adjusting the data transmission path, limiting the specific type of traffic, etc.

3. Integration and training: Integrate the designed neural network model into the computer network as part of the intelligent operator. Reinforcement learning algorithms, such as the experiential playback mechanism of deep Q networks, are then used to train neural network models from data collected in real time. During the training process, the neural network will constantly interact with the network, learn and adjust according to the actual network state and feedback signals, in order to improve its network traffic management ability.

4. Online deployment and real-time adjustment: After the training is completed, the trained neural network model is deployed to the actual network environment. The intelligent operator will monitor network traffic in real time and make operational decisions based on the output of the neural network model. At the same time, on-the-ground data can be continuously collected for further optimization and adaptation of the neural network model to the changing network environment and needs[10].

Through this integrated approach, we achieve a close combination of computer networks and neural networks, enabling intelligent operators to collect and train data from the network in real time, thereby continuously improving their network management and optimization capabilities. This integrated approach not only improves the intelligence level of network management, but also brings higher efficiency and reliability to network operation and maintenance.

*2.4. Work Overview*

The main goal of this paper is to create an intelligent network operator capable of performing general-purpose computer network analysis tasks to solve a wide variety of problems. In order to achieve this goal, we have carried out the following work:

First, we designed an architecture that integrates computer networks and neural networks. This architecture can effectively combine the knowledge of computer network with the learning ability of neural network, so as to realize the processing and analysis of network tasks by intelligent operators.

Second, we implemented this design architecture and successfully trained the intelligent operator. Through a large amount of training data and optimization algorithms, we enable intelligent operators with high analytical power and accuracy. At the same time, we also made the source code of the project available so that other researchers can further explore and apply our results.

Finally, we propose a critical loss algorithm that enables an intelligent operator to achieve 100% accuracy. By optimizing and adjusting the loss function, we can effectively improve the performance of intelligent operators in network analysis tasks, and provide reliable support and guarantee for solving practical network problems.

In conclusion, by integrating computer networks and neural networks together, we are able to enable the training and deployment of intelligent operators, thereby automating and intelligentizing network management. Intelligent operators can monitor network traffic in real time, analyze packets, and make decisions based on pre-designed neural network models to optimize network performance and management efficiency. This integrated approach not only improves the level of intelligence in network management, but also brings greater efficiency and reliability to network operations and maintenance.

On the basis of the Python open source project stable-baselines3, the code of the project is implemented to achieve the design function of intelligent operators, and the complete code is stored on GitHub to become a new open source project, hoping that more engineers can participate in it and jointly improve the intelligence level of the industry.

The project site is: https://github.com/Andy-Wu-2022/ComputerNetworkOperator.

In the next step, we can further optimize and adjust the neural network model to adapt to the changing network environment and needs. At the same time, we can also explore more network management tasks and scenarios, such as security protection, resource scheduling, etc., to further improve the functions and performance of intelligent operators. Through continuous research and practice, we will continue to promote the development of network management technology, and provide better support and services for the construction of intelligent and efficient network systems.

**3. Methodology**

This paper mainly uses four key modules to realize the designed architecture, including computer network simulator, computer network environment component, trainer and tester. Firstly, the computer network simulator provides a simulation environment for the system, increases the diversity of data by randomly creating the computer network, and provides a rich data set for the training of intelligent operators. Second, the computer network environment component acts as a bridge connecting the simulator and the intelligent operator, responsible for collecting data in real time and passing it to the operator for processing and decision making. [11-12]The trainer module then uses the collected data and a pre-designed neural network model to train the intelligent operator through reinforcement learning algorithms to continuously optimize its strategy and behavior. Finally, the tester module evaluates the trained intelligent operator's performance, testing and validating its behavior through real or simulated network environments to evaluate its performance in different situations. These four modules together build a complete intelligent operator system, realize the intelligent management and optimization of computer network, and provide important support for the construction of intelligent and efficient network system.

*3.1. Computer network simulator*

The simulator has the ability to randomly generate simple computer networks, which allows users to easily create networks with different topologies, thus increasing the flexibility and diversity of the system (Figure 2). Users can update the network status as required, including node connection, device working status, and network traffic load. In addition, the simulator also provides several other functions, such as real-time monitoring of network performance, simulating network failures and attacks, and evaluating network security. Through these functions, users can fully understand and evaluate the operation of the simulated network, so as to better understand the working principle and characteristics of the computer network

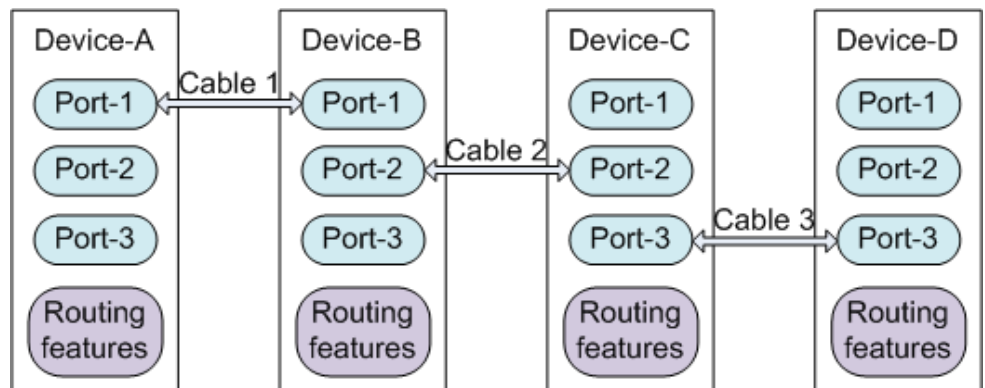

**Figure 2.** Computer network diagram

The computer network generated by the emulator shown in Figure 2 has the following characteristics:
**a.** The number of devices ranges from 4 to 10.
**b.** Devices are connected through the tree topology.
**c.** IP Subnet Select a subnet from a pool containing 32 subnets. For brevity, the subnets are named IP-Subnet-1 through IP-Subnet-32. The mask of a subnet is a 24-bit 255.255.255.0.
**d.** Select an IP address from a pool of 63 IP addresses corresponding to each subnet. For brevity, IP addresses are named from ip-address -1 to ip-address -63. The full name of an IP Address is "ip-address -x in IP- subnet-y" (x and y are numbers). In this article, only "ip-address -x" is used to indicate the name of the IP Address, which is more concise and still valid because the IP Subnet to which the IP address belongs is fixed and immutable.
**e.** Routing Protocol Select RIP version 2, EIGRP (AS 1), and OSPF (Process ID 1). The automatic summary function of RIP and EIGRP is disabled. The area of all OSPF subnets is 0.
**f.** In this example network, the following six types of faults are randomly inserted:

**Table 1.**Device Status and Possible Reasons

| Status | Possible Reasons |
|---|---|
| Port Closed | 1. Port misconfiguration or not properly enabled. |
| | 2. Port not specified in device configuration. |
| | 3. Hardware fault or connectivity issues. |
| Incorrect IP Address | 1. Static IP address misconfigured. |
| | 2. Dynamic IP address allocation failure or misconfiguration. |
| | 3. Subnet mask misconfigured. |
| Incorrect IP Subnet | 1. Subnet mask misconfigured. |
| | 2. IP address not matching the network of the device. |
| | 3. Subnet mask not matching the network of the device. |
| Missing IP Subnet | 1. Device not configured with an IP address. |
| | 2. Device not joined to any network. |
| | 3. Non-existent network or connectivity issues. |
| Automatic Summary Routing | 1. Automatic summary function misconfigured. |
| | 2. Router not enabled with automatic summary function. |
| | 3. Erroneous summary routes in the static routing table. |
| Routing Protocol Version | 1. Routing protocol misconfiguration. |
| | 2. Protocol version mismatch between router and neighboring devices. |
| | 3. Router not enabled with any routing protocol. |

The above features are very important for any type of computer network, and more functions can be added in future work, such as: security functions, quality of service functions, traffic engineering functions, etc.

*3.2. Computer network environment components*

The environment component inherits all the functionality of the simulator and adds two additional features: an encoder that converts the observation information text into numbers (the observation information encoder) and an encoder that converts the observation information into embedded vectors (the observation information embedder).

The observation information embedder embeds numeric information into an embedded vector consisting of floating point numbers (vectors). This article uses numbers between 0 and 1 because the relationships between network information used in the environment components are relatively simple. The principle is to bring the embedding vectors to which the same kind of numeric information belongs close to each other and distance them from the embedding vectors of other classes, so that adjacent numbers will have some common feature or meaning. For example, IP addresses are classified as category 1, and during training, operators will be able to learn this feature.

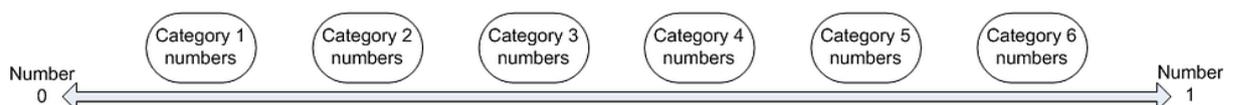

**Figure 3.** Concept of embedding vectors

At each step, the environment component uses a portion of the computer network information (the corresponding information obtained from the initial network information dictionary and the current network information dictionary) as observation information, such as link state information for a port, or IP address information for a port. It loops through all network information until all problems are fixed or the maximum number of allowed steps is reached.

The environment component also breaks down the task of generating complex instructions into substeps. For example, in order to configure the correct IP address, there are 3 sub-steps:

**a.** The operator calculates the observation information and outputs an instruction indicating that there is a problem with the port.

**b.** The operator calculates the observation information and outputs a command indicating the correct command.

**c.** The operator calculates the observation information and outputs an instruction, representing the parameters of the instruction.

The environment component then combines these three subinstructions into a complete network device instruction for further processing.

*3.3. Practical application*

In practice and application, the simulator randomly generates A simple computer network by following these steps: First, randomly generates the number of devices (between 4 and 10), assigning each Device a name, such as Device-a. Then, randomly select one device as the root device and connect another device to the root device using cable 1. Subsequently, more devices are connected to the existing network using one cable at a time until there are no devices left. Next, randomly select a routing protocol from three options: RIP, EIGRP, or OSPF. IP addresses and IP subnets are then randomly selected from a predefined pool and the information is assigned to each device. Generate initial network information, including a description of the network design and an initial network information dictionary. At this point, there is no fault in the network, and all parts are in the correct state of design. Randomly generate a list of faults to inject into the network and inject the faults into the network, generating a dictionary of current network information with correct or incorrect states. In addition, the simulator implements a simplified version of the network estimator, a computer network environment component that calculates rewards by checking that the operator's action instructions are in the correct instruction list. If the operation instructions are correct, the reward is set to 1; If the operation instruction is wrong, the reward is set to -1.

**4. Critical Loss Algorithm**

*4.1. Model training*

The intelligent operator (neural network model) is randomly initialized at the beginning, at which point it can only randomly generate instructions. During training, the operator gradually learns the characteristics of the input data (observation information) and begins to generate more and more correct instructions. The process is described in detail as follows:

a. The environment component reads a portion of the initial network information dictionary and the current network information dictionary (for example, IP address information for a port) as text observations, encodes and emphases the observations into an embedding vector (current observations), and then feeds it to the operator for analysis.

b. The operator calculates the embedded vector and outputs an instruction (current instruction).

c. The environment component decodes the instructions into network device instructions, applies them to the computer network, then evaluates new network information and issues rewards (current rewards. Correct instructions use the number 1, wrong instructions use the number -1). At the same time, the environment component generates the next embedding vector (the next observation information).

d. The data set collects and stores the current observation information, the current instruction, the current reward, and the next observation information as a sample in its buffer.

e. Repeat the above steps to collect a large number of samples into the data set. Some samples are positive (instructions are correct), others are negative (instructions are wrong).

f. Use algorithms and datasets to train operators.

g. The above steps can be performed multiple times until the operator is able to generate 100% correct instructions (see the introduction documentation on GitHub for more details on training if needed).

*4.2. Critical loss function*

The critical loss algorithm proposed in this paper is improved on the basis of QRDQN algorithm.

QRDQN algorithm is one of the most mainstream reinforcement learning algorithms at present, it will calculate each state-action Pair and get multiple evaluation scores, and finally select the Action with the highest average score as the correct instruction under the State and output the instruction.

The accuracy of the original QRDQN algorithm is around 80%, which means that about 20% of the output instructions are wrong. This is due to the following two characteristics of the neural network itself:

1. The neural network will output calculated data in the statistical sense, whose value represents the correct probability of a certain result. The higher the value, the greater the probability that the result is correct.

2. In the optimization process of the neural network, its calculation parameters will be fine-tuned, which brings the volatility of the output value and affects its accuracy.

*4.3. solution*

1. Set the target value of the output value according to the demand. In this project, we need to separate the correct and wrong instructions, so we need to select two target values. On the one hand, the selected values need to be between 0 and 1 (the data sensitive interval of neural network is -1 to 1, because only two target values are needed, so choose two positive values to reduce the computational complexity), on the other hand, there should be appropriate spacing between them to effectively separate the right and wrong instructions. It is also necessary to stay away from the two boundaries of 0 and 1 (because of the problem of precision degradation at the boundary, which requires complex numerical processing methods). Finally, 0.7 is selected as the target value of the output value of the correct instruction, and 0.3 is selected as the target value of the output value of the wrong instruction.

2. The error between the output value and the target value of the neural network is called loss, and some losses will lead to errors in the instruction optimization process, and they are classified as critical losses; Other losses have less impact on the instruction selection process and are classified as non-critical losses. In the calculation of total loss and optimization of neural network, the critical loss is given a larger weight, and the key loss is reduced and the accuracy is improved in the optimization process.

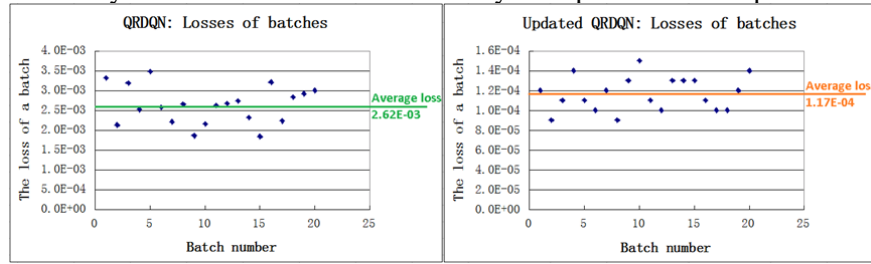

**Figure 4.** Ratio before and after critical loss

In the process of algorithm optimization, a neural network is configured to output 7 numerical values ranging between 0 and 1, representing the evaluation scores of instructions. The average of these scores determines the quality of an instruction, with higher averages indicating better instructions. This approach mitigates the impact of neural network output volatility on accuracy rates. Optimal instructions are selected based on these evaluations.

Following the application of optimal instructions to the computer network, their effectiveness is evaluated. Instructions are categorized as correct or incorrect based on their outcomes.

To calculate critical losses, four values are considered:

The target evaluation score for correct instructions is set at 0.7. Scores equal to or greater than 0.7 are associated with non-critical losses.

The safety threshold for the average evaluation score of correct instructions is 0.56. An average score equal to or greater than this value is also associated with non-critical losses.

The target evaluation score for incorrect instructions is set at 0.3. Scores equal to or less than 0.3 are associated with non-critical losses.

The safety threshold for the average evaluation score of incorrect instructions is 0.44. An average score less than or equal to this value is associated with non-critical losses.

All other losses are classified as material losses.

The Adam optimization algorithm is employed to optimize the neural network based on critical loss analysis. This optimization leads to a significant reduction in average loss, from 2.62E-03 to 1.17E-04, representing a 95.5% decrease.

Further fine-tuning results in a 100% accuracy rate for output instructions(figure 4).

## 5. Conclusion

The performance of the intelligent operator has been successfully demonstrated through extensive testing on over 12,000 randomly generated computer networks, as detailed in Section 2.4 and documented on the GitHub repository provided. The intelligent operator, implemented as a neural network model, exhibits remarkable accuracy, achieving 100% accuracy across 28 issues and 237 operations per network, totaling over 2.8 million operations without error. Detailed output from these tests is available in the project documentation on GitHub. With processing times ranging from 1 to 5 seconds per network inspection and repair, the intelligent operator offers efficient solutions to a diverse range of computer network issues. It is noteworthy that initial loading of the operator takes approximately 10 seconds, and printing detailed information for all steps requires an additional 10 seconds[13-14].

In conclusion, this paper has successfully realized an AI-powered network operator, demonstrating its efficacy in efficiently addressing a variety of computer network issues. The integration of intelligent operators with computer networks and neural networks presents a promising and challenging avenue for future research and application in network management and optimization.